\documentclass[aps,pre,amssymb,superscriptaddress,showpacs,floatfix,a4paper,
twocolumn,10pt]{revtex4-1}

\usepackage{graphicx}
\usepackage{bm}
\usepackage{amsmath}
\usepackage{epsf}
\usepackage{epsfig}
\usepackage{amssymb}

\begin{document}

\title{Thermal rounding exponent of the depinning transition of an elastic
string in a random medium}

\author{S. Bustingorry}
\affiliation{CONICET, Centro At{\'{o}}mico Bariloche, 8400 San Carlos de
Bariloche, R\'{\i}o Negro, Argentina}
\author{A. B. Kolton}
\affiliation{CONICET, Centro At{\'{o}}mico Bariloche, 8400 San Carlos de
Bariloche, R\'{\i}o Negro, Argentina}
\author{T. Giamarchi}
\affiliation{DPMC-MaNEP, University of Geneva, 24 Quai Ernest Ansermet, 1211
Geneva 4, Switzerland}


\begin{abstract}
We study numerically thermal effects at the depinning transition of an elastic
string driven in a two-dimensional uncorrelated disorder potential. The velocity of the string
exactly at the sample critical force is shown to behave as $V \sim T^\psi$, with
$\psi$ the thermal rounding exponent. We show that the computed value of the
thermal rounding exponent, $\psi = 0.15$, is robust and accounts for the different scaling
properties of several observables both in the steady-state and in the transient
relaxation to the steady-state. In particular, we show the compatibility of the
thermal rounding exponent with the scaling properties of the steady-state
structure factor, the universal short-time dynamics of the transient velocity at
the sample critical force, and the velocity scaling function describing the
joint dependence of the steady-state velocity on the external drive and
temperature.
\end{abstract}


\maketitle

\section{Introduction}
\label{sec:intro}

The understanding of the static and dynamic properties of elastic interfaces in
disordered media has direct impact on different fields on condensed matter
physics. Among a large variety of systems one can mention
magnetic~\cite{lemerle_domainwall_creep,bauer_deroughening_magnetic2,
yamanouchi_creep_ferromagnetic_semiconductor2,metaxas_depinning_thermal_rounding}
or ferroelectric~\cite{paruch_ferro_roughness_dipolar,paruch_ferro_quench}
domain walls, contact lines~\cite{moulinet_distribution_width_contact_line2},
fractures~\cite{bouchaud_crack_propagation2,alava_review_cracks}, vortex
lattices~\cite{blatter_vortex_review,giamarchi_vortex_review,
du_aging_bragg_glass}, charge density waves~\cite{nattermann_cdw_review}, and
Wigner crystals~\cite{giamarchi_electronic_crystals_review}, as paradigmatic
examples. Since the effect of the disordered media in all these systems is
non-trivial, an important question is how these elastic objects respond to an
external drive.

When the temperature is zero, there exists a critical force value $F_c$ such
that the steady state velocity of the center of mass of the interface is zero
below $F_c$ and is finite above it. This is due to the complex interplay between
disorder and external force: the interface accommodates within the disorder
energy landscape and a finite energy barrier must be overcome by the external
force in order to generate a net movement. Therefore a finite force value has to
be set to have an infinitesimally small finite velocity. This is the so
called depinning transition. If the critical force value is approached from
above, the velocity vanishes as $V \sim (F-F_c)^\beta$ for a thermodynamic
system, with $\beta$ the depinning exponent. Concomitant with the power-law
decrease of the velocity is the divergence of a characteristic length as $\xi
\sim (F-F_c)^{-\nu}$, with $\nu$ the correlation length exponent. This depinning
correlation length gives the typical size of the correlated displacement (or
avalanche) that makes the interface advance in the direction of the external
force. The finite force threshold, the critical decrease of the velocity order
parameter and the divergence of the typical length scale led to propose a
description of the depinning transition using tools from standard critical
phenomena~\cite{fisher_depinning_meanfield}. More recently however, the 
analysis of the low-temperature averaged steady-state geometry has shown that no
divergent steady-state correlation length-scale exists approaching the critical
force from below, thus breaking the naive analogy with standard phase
transitions, 
where two divergent length-scales are expected above and below the critical
point~\cite{kolton_depinning_zerot2,kolton_creep_exact_pathways}.

When the temperature is finite there is no sharp transition between zero and
finite velocity regimes. Even at forces much smaller that the critical value the
interface is able to move since thermal activation is enough to overcome the
effective energy barriers generated by the disorder. This regime, $F \ll F_c$,
is the creep regime, and it is characterized by a stretched exponential
dependence of the velocity with the inverse of the external
force~\cite{ioffe_creep,nattermann_creep,feigelman_creep,nattermann_creep_law,
chauve_creep_short,chauve_creep_long}. On the other hand, at forces around the
critical value, $F \approx F_c$, a finite temperature value smears out the
transition, which is no longer abrupt. This thermal rounding of the depinning
transition can be characterized, exactly at the critical force $F=F_c$, by a
power-law vanishing of the velocity with the temperature as $V \sim T^\psi$,
with $\psi$ the thermal rounding
exponent~\cite{middleton_CDW_thermal_exponent,chen_marchetti,
nowak_thermal_rounding,roters_thermal_rounding1,
vandembroucq_thermal_rounding_extremal_model,luo_thermal_rounding_flux_lines,
bustingorry_thermal_rounding_epl}.

The values of the different exponents characterizing the depinning transition
are universal in the sense that their values depend on few parameters of the
system such as the range of the intrinsic elasticity, the dimensionality of the
problem, and the correlated structure of the disorder. For the experimentally
relevant case of $1+1$ dimensional elastic interfaces moving in a random-bond
disorder environment with short-range correlations and short-range elasticity,
we have recently reported the value $\psi = 0.15 \pm 0.01$ using Langevin
dynamics numerical simulations~\cite{bustingorry_thermal_rounding_epl}. This
value compares well with the value $\psi = 0.16$ reported in
Ref.~\cite{chen_marchetti} based in numerical simulations. However, these values
are smaller than the value $\psi=0.24$ obtained using an artificial extremal
activated dynamics~\cite{vandembroucq_thermal_rounding_extremal_model}, which
might indeed be in a different universality class. The value $\psi = 0.2$ was
obtained using numerical simulations of domain wall motion with the random-field
Ising model~\cite{nowak_thermal_rounding,roters_thermal_rounding1}. Although it
is expected that for $T>0$ and around the depinning transition the
characteristic exponents do not depend on the random-bond or random-field
character of the disorder, this slightly larger value might be possibly ascribed
to the anharmonic corrections to the elasticity present in the random-field
Ising model. On the other hand, functional renormalization group equations at
the depinning~\cite{chauve_creep_long} allow in principle to extract the thermal
rounding exponents. However, in practice there are, up to now, no analytical
estimates of $\psi$, unlike the other critical exponents which have been
computed using functional renormalization group up to two
loops~\cite{ledoussal_frg_twoloops}. The very existence of a
thermal rounding, obeying a power law scaling is not rigorously proven, and
there are indeed some models of depinning which exhibit at finite temperature a
totally different type of thermal rounding~\cite{lecomte_wires_depinning}. It is
thus crucial, given the uncertainty on the very type of thermal rounding and
certainly on the value of the thermal rounding exponent, to develop new methods
to determine $\psi$, and to check the robustness, consistency, and expected
universality of the phenomenological scaling arguments.

Experimentally, access to the full force range relevant to the depinning
transition has been reported in ultrathin ferromagnetic
layers~\cite{metaxas_depinning_thermal_rounding,metaxas_thesis,
metaxas_coupled_interfaces}. In this case, the thermal rounding of the depinning
transition is generated through an effective temperature dependence controlled
by the relative disorder intensity among different samples. Indeed, it has been
shown that thermal effects on the velocity-force characteristics can be well
described using the value $\psi=0.15 \pm 0.10$~\cite{metaxas_thesis}.

The aim of the present work is to give further numerical support to the reported
value $\psi=0.15$, by checking the robustness and consistency of the scaling
arguments applied to different observables. To this end, we show how this value
allows to describe different measures characterizing the critical behavior of
the depinning transition: an analysis of the finite temperature structure
factor, a short-time dynamics analysis, and the analysis of the scaling function
describing the velocity dependence on force and temperature around depinning for
different disorder intensities.

\section{Model system and numerical simulations}
\label{sec:model}

In order to model the dynamics of one-dimensional interfaces in disordered media
we use a short-range elastic string, as described in the following. The string
is defined by a single valued function $u(z,t)$, giving its transverse position
$u$ in the $z$ axis. The time evolution of the string is given by the overdamped
equation of motion
\begin{equation}
\label{eq:EW-F}
\gamma \, \partial_t u(z,t) = c \, \partial^2_z u(z,t) + F_p(u,z) + F +
\eta(z,t),
\end{equation}
where $\gamma$ is the friction coefficient and $c$ the elastic constant. The
pinning force comes from the derivative of the random-bond pinning potential
$U(u,z)$, i.e. $F_p(u,z) = - \partial_u U(u,z)$, whose sample to sample
fluctuations are given by
\begin{equation}
 \overline{\left[ U(u,z)- U(u',z') \right]^2} = \delta(z-z') \, R^2(u-u')
\end{equation}
where $R(u)$ stands for a correlator of range $r_f$~\cite{chauve_creep_short},
and the overline indicates average over disorder realizations. Thermal
fluctuations are included through the thermal noise term $\eta(z,t)$ which
satisfies
\begin{eqnarray}
 \langle \eta(z,t) \rangle &=&0, \nonumber \\ 
 \langle \eta(z,t) \eta(z',t') \rangle &=& 2 \gamma T \delta(t-t') \delta(u-u'),
\end{eqnarray}
where $T$ is the temperature (with Boltzmann constant set to unity, $k_B = 1$)
and the angular brackets denote a thermal average. Finally, the force $F$ on
Eq.~\eqref{eq:EW-F} corresponds to a uniform and constant external field which
drives the string in the $u$ direction.

The evolution Eq.~\eqref{eq:EW-F} is numerically solved. The $z$ direction is
discretized in $L$ segments of size $\delta z=1$, i.e. $z \to j=0,...,L-1$,
while keeping $u_j(t)$ as a continuous variable. This sets the longitudinal
finite system size $L$. The equation is integrated using the Euler method with a
time step $\delta t=0.01$. The pinning potential is modeled by performing
a cubic spline passing through $M$ regularly spaced uncorrelated
Gaussian numbers points~\cite{rosso_depinning_simulation,kolton_creep2}, which
sets the transverse finite system size $M$. Numerical simulations are performed
using periodic boundary conditions in both directions and using the parameters
$\gamma=1$, $c=1$, and $r_f=1$. The strength of the disorder is given by
$R_0=R(0)$. For each disorder realization, i.e. for each finite size sample, the
critical force $F_c$ can be accurately obtained using an exact algorithm, which
also gives the critical pinned configuration of the string
$u_c(z)$~\cite{rosso_depinning_simulation}. The results presented in the
following sections were obtained by typically averaging over $100$ disorder
configurations; the error bars being typically of the order of the size of the
data points.

\section{Velocity-force characteristics: scaled variables}
\label{sec:vdef}

\begin{figure}[!tbp]
\includegraphics[angle=-0,width=7.5cm,clip=true]{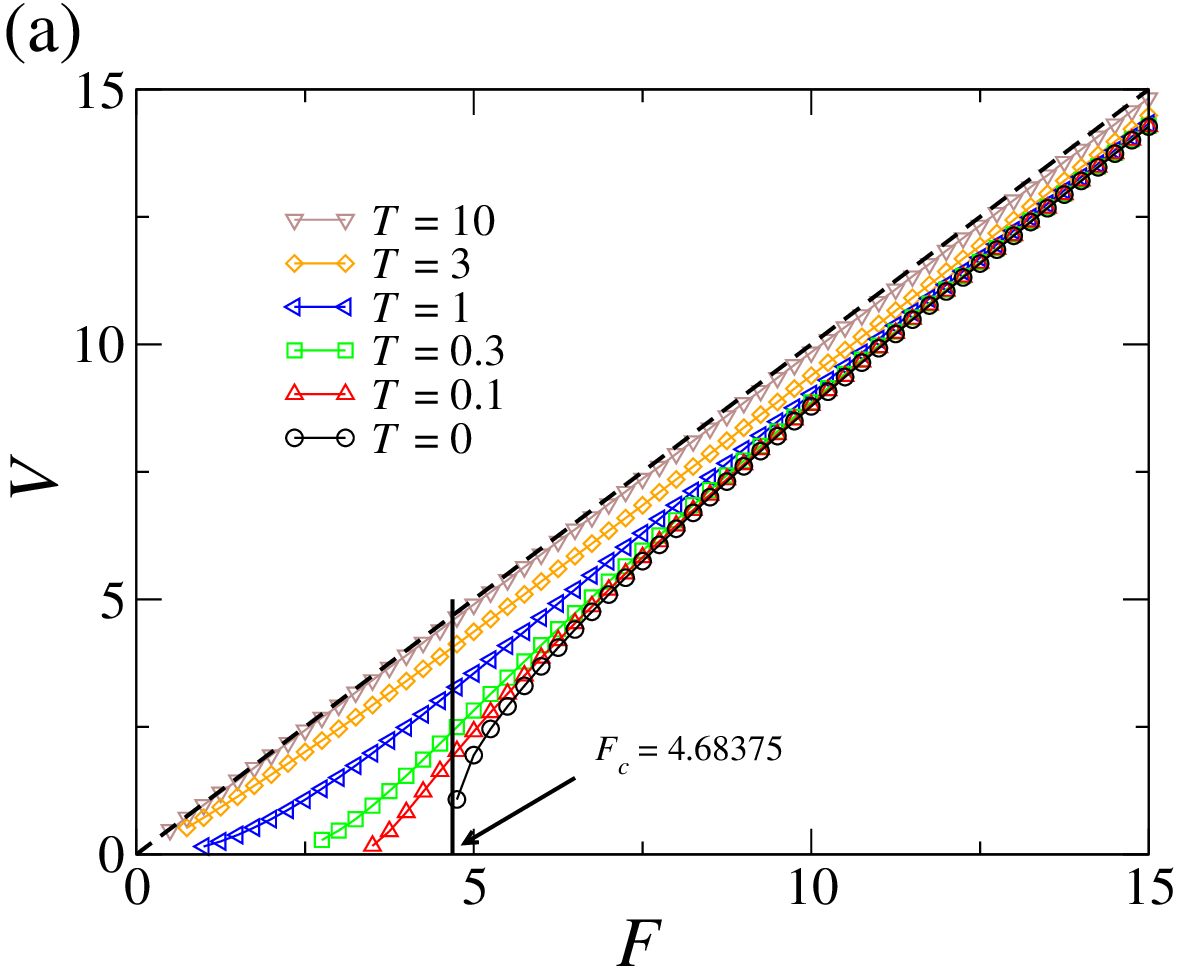}
\includegraphics[angle=-0,width=7.5cm,clip=true]{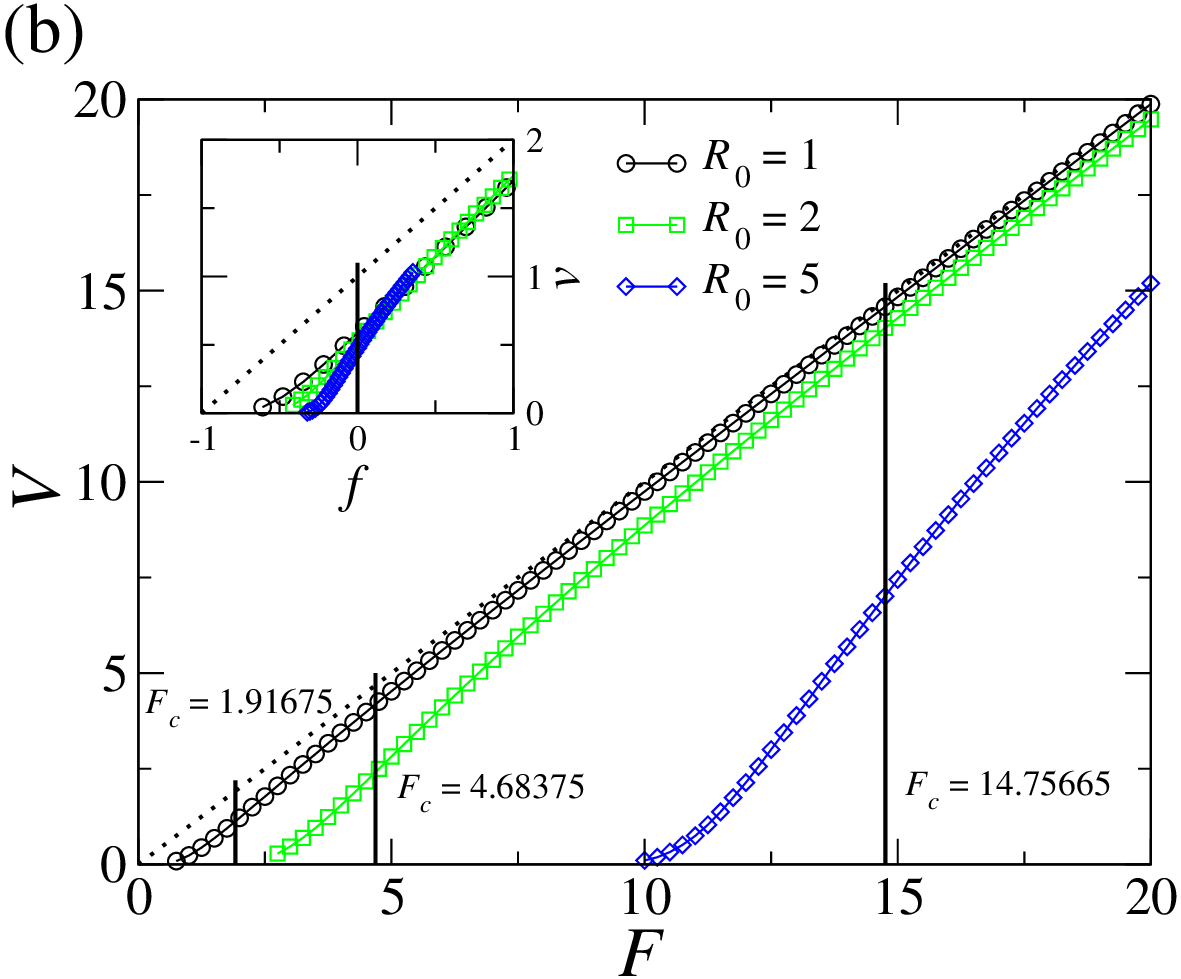}
\caption{\label{f:vdeFyT} (Color online) (a) Velocity-force characteristics for a single
disorder realization of intensity $R_0=2$ and different temperatures. (b)
Velocity-force characteristic for temperature $T=0.3$ and different disorder
intensities as indicated. The inset shows the scaled data according to $v=V/(m F_c)$
and $f=(F-F_c)/F_c$, where $F_c$ is the sample dependent critical force.
Furthermore, this rescaling strongly reduces sample-to-sample fluctuations.}
\end{figure}

In this section we will present the general features of the velocity-force
characteristics, allowing us to define the critical region and the scaled
variables that will be used throughout the rest of this work.
Figure~\ref{f:vdeFyT}(a) shows typical velocity-force curves at finite
temperature, as obtained with the present model 
for $L=1024$ and $M = 5792 \approx L^{\zeta_{\mathrm{dep}}}$, 
with $\zeta_{\mathrm{dep}}=1.25$~\cite{duemmer2,rosso_hartmann} the depinning
roughness exponent (see below).
Given a fixed force $F$, the
velocity is computed in the steady state, which is typically reached within one
sweep over the transverse size $M$ (as detailed below, the transverse size $M$ 
will be varied following a scaling relation with the string length $L$).
Then, of the order of five sweeps over $M$ are used to compute the velocity
\begin{equation}
 V = \langle \partial_t u(z,t) \rangle.
\end{equation}
The thermal average is taken by computing $200$ values of the velocity with
independent thermal noise realizations within this steady state regime.
Different curves correspond to the same \textit{single} disorder realization
with intensity $R_0=2$ and increasing temperature. The characteristic critical
force is indicated in the key. The lower curve, corresponding to $T=0$, clearly
presents the typical abrupt depinning transition: the velocity is strictly zero
for $F < F_c$, while it increases as $(F - F_c)^\beta$ for $F > F_c$, where
$\beta < 1$ is the velocity exponent. As observed, by increasing the temperature
the $T=0$ sharp transition is smeared out. Although at very small temperatures
the curves still present the curvature corresponding to $(F - F_c)^\beta$, at
higher temperatures there are no clear signatures of the underlying $T=0$
depinning transition. Finally, at very high temperatures, when the thermal
energy is larger than the typical pinning energy, the velocity tends to
increase linearly with the force, $V = mF$, with the mobility $m=1/\gamma$,
corresponding to the dashed line in Fig.~\ref{f:vdeFyT}(a).

In Fig.~\ref{f:vdeFyT}(b) the disorder intensity effects on the velocity-force
characteristic can be observed. The critical forces for the corresponding
disorder realizations for each intensity are quoted. Since around the depinning
transition the velocity strongly depends on the sample critical force value,
along the present work we will use scaled variables for velocity and force. The
scaled velocity is given by
\begin{equation}
\label{eq:vscal}
 v = \overline{\left(\frac{\langle \partial_t u(z,t) \rangle}{m F_c}\right)} =
\overline{\left(\frac{V}{m F_c}\right)}, 
\end{equation}
which defines a systematics to average over disorder realizations. Besides, we
use as the control parameter the scaled force
\begin{equation}
\label{eq:fscal}
 f = \frac{F-F_c}{F_c},
\end{equation}
which measures the scaled distance to the critical force for each disorder
realization. These definitions of scaled variables are different than the scaled
variables used in standard critical phenomena. In our case, we are using the
critical force of \textit{each} disorder realization in order to measure how
close the system is to the critical point, instead of using the disorder
\textit{averaged} value $\overline{F_c}$. Besides, we also incorporate into the
definition of the order parameter $v$ a non-trivial disorder average when using
the disorder realization dependent value $F_c$. 

Although a temperature-dependent critical force can be considered
for studying thermal properties at depinning~\cite{blatter_vortex_review}, we are using here the 
zero-temperature value. From a practical point of view the temperature-dependent
critical force can be defined as the inflexion point of the velocity-force curves.
In fact, in the temperature range we are studying here this temperature-dependent critical
force does not deviate much from the zero-temperature value.
Instead of using a temperature-dependent critical force, we adhere here
to the idea that the important quantity given by the disorder potential is the
zero-temperature critical force and that the small temperature data can be interpreted
using this quantity. Besides, the zero-temperature critical force strictly depends only
on the disorder configuration and therefore permits the computation of the 
average velocity using Eq.~\eqref{eq:vscal} with disorder and temperature averages
independently realized.

The scaled variables, Eqs.~\eqref{eq:vscal} and \eqref{eq:fscal},
are natural for an overdamped particle driven in a periodic
potential $U(x)=R_0 \cos(x/\lambda)$, $\gamma dx/dt = -dU(x)/dx + F$, where one can readily obtain 
that $F_c = R_0/\lambda$ and $\gamma V/F_c \approx \sqrt{(F-F_c)/F_c}$ close to $F_c$.
They also arise in functional renormalization group calculations for  
center of mass velocity of an elastic manifold, $\tilde\gamma  V/F_c \sim [(F-F_c)/F_c]^\beta$
but with $\tilde \gamma$ an effective friction coefficient~\cite{chauve_creep_long}. 
On the other hand, from a practical point of view, it was shown that the scaled variables 
defined in Eqs.~\eqref{eq:vscal} and \eqref{eq:fscal}, applied to each particular sample, 
serve to diminish sample-to-sample 
fluctuations when studying the depinning transition of a
string~\cite{bustingorry_thermal_rounding_epl,duemmer2}. 

The inset of Fig.~\ref{f:vdeFyT}(b) shows the same data as in the main panel but in scaled form 
for a single realization. The difference between these curves close to the threshold 
is due to the fact that the full function $V(F,T)$ depends on the disorder intensity not only
through the value of the critical force, but one also needs to consider both extra
disorder-dependent temperature scale and friction coefficient. This 
interesting issue will not be crucial for our present study however (see 
discussion in Sec.~\ref{sec:univ-func} below).

\section{Temperature dependence of the velocity at the critical force}
\label{sec:vdeT}

\begin{figure}[!tbp]
\includegraphics[angle=-0,width=7.5cm,clip=true]{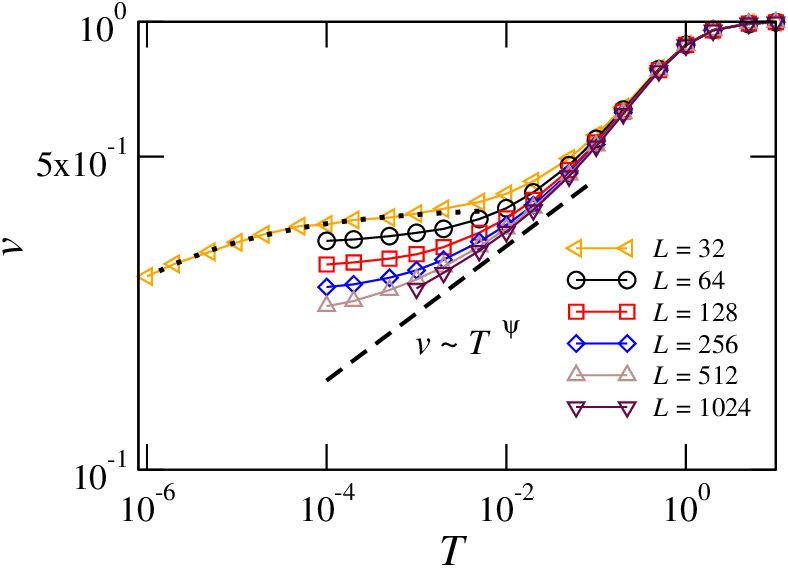}
\caption{\label{f:vdeT-scal-Ap05-g-a} (Color online)
Velocity-temperature curves for different
systems sizes $L$, as indicated, while keeping $M=4\,L^{\zeta_{\mathrm{dep}}}$,
with $\zeta_{\mathrm{dep}}=1.25$~\cite{duemmer2,rosso_hartmann} the depinning
roughness exponent. The disorder intensity is $R_0 = 0.5$. All the data were
computed at exactly the sample critical force and then averaged over disorder
realizations. The dashed line corresponds to the expected power-law behavior.
The dotted line describes the crossover to the one-particle regime as discussed
in the text.}
\end{figure}

Here we show the finite temperature response of the elastic string exactly at
the critical force and discuss some finite size scaling effects, in particular
the crossover to single particle dynamics. Figure~\ref{f:vdeT-scal-Ap05-g-a}
presents velocity-temperature curves for different system sizes. All the curves
were computed at exactly the sample critical force, using the scaled force
variable $f=0$, and then averaged over disorder realizations. The disorder
intensity is $R_0 = 0.5$ and the results are qualitatively similar to those
reported for $R_0 = 1$ in Ref.~\cite{bustingorry_thermal_rounding_epl}. At very
high temperatures, $T \gg R_0$, the system enters the fast flow regime and the
velocity practically equals the force; therefore the reduced velocity (which
incorporates the critical force) tends to unity. At intermediate temperatures,
the velocity reduces and the curves tend to display the critical behavior $v
\sim T^\psi$. This power-law behavior is however interrupted by finite-size
effects at smaller temperatures, when the dynamic characteristic length $\xi$ 
equals the system size $L$.

At very small temperatures a crossover to single particle
dynamics~\cite{duemmer2,kolton_universal_aging_at_depinning} is observed as
shown by the $L=32$ curve. A simple ad-hoc model to rationalize this crossover
has been given by Duemmer and Krauth~\cite{duemmer2} while numerically studying
the zero-temperature depinning transition. Within this model, one can write the
velocity in the very small temperature regime as
\begin{equation}
\label{eq:vfitlowT}
 v = \frac{M}{\tau_0+\tau_1(T)},
\end{equation}
where $\tau_1(T)$ is the temperature-dependent time the interface spends near
the critical configuration and $\tau_0$ is the rest of the time needed to cover
the transverse spatial period $M$ of the computational box. In this simple
model, $\tau_0$ is approximated to be temperature independent at very low
temperatures. Using the temperature dependence of the escape rate for a particle
in a random potential~\cite{colet_marginal_thermal_escape}, one can propose that
$\tau_1 = a T^{-1/3}$. In Fig.~\ref{f:vdeT-scal-Ap05-g-a} we show with a dotted
line that the very small temperature regime for $L=32$ is well fitted with
Eq.~\eqref{eq:vfitlowT}. For $L=32$ and $M=304 \approx 4L^{\zeta_{\mathrm{dep}}}$
we fitted the $T<10^{-3}$ regime using Eq.~\eqref{eq:vfitlowT} and we found the
fitting parameters $\tau_0=779.5$ and $a=3.42$. This is a simplified model
allowing to rationalize the crossover to one-particle dynamics and should be
further tested.

\begin{figure}[!tbp]
\includegraphics[angle=-0,width=8cm,clip=true]{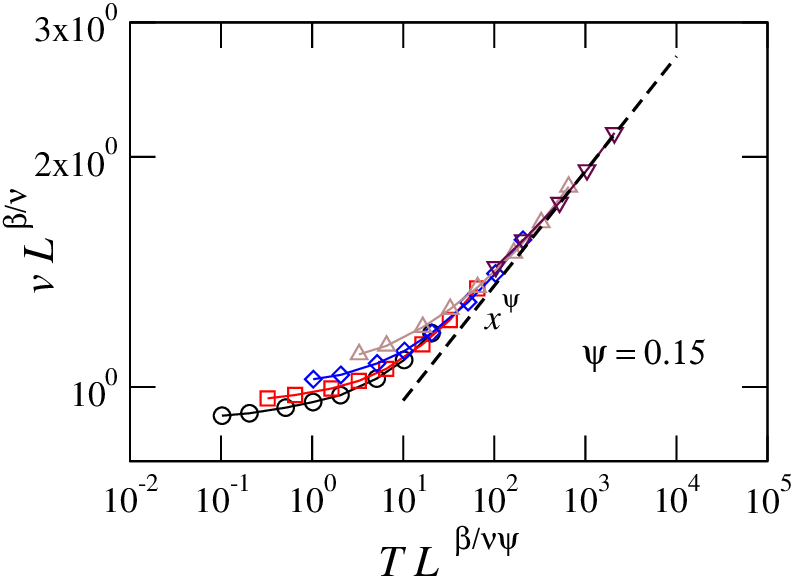}
\caption{\label{f:vdeT-scal-Ap05-n} (Color online) Finite size scaling of the
velocity-temperature curves for different $L$ values according to
Eq.~\eqref{eq:std-fsc}. The disorder strength is $R_0 = 0.5$ and the transverse
size has been kept at $M=4\,L^{\zeta_{\mathrm{dep}}}$. We show points for $T \le
0.02$ and $L \ge 64$, since for $L=32$ the single particle regime is present at
small temperatures. The dashed line corresponds to the power-law behavior with
$\psi = 0.15$. We also used the values $\beta=0.33$~\cite{duemmer2} and
$\nu=1.33$~\cite{kolton_short_time_exponents}.}
\end{figure}

The finite size effects displayed by the velocity-temperature curves in
Fig.~\ref{f:vdeT-scal-Ap05-g-a} are not easily accounted for by standard finite
size scaling arguments. In fact, assuming finite-size scaling as in standard
critical phenomena, the velocity should be described by universal functions as in
\begin{equation}
\label{eq:std-fsc}
 v = L^{-\beta/\nu} g\left(L^{\beta/\nu\psi}\, T \right),
\end{equation}
with $g(x) \sim 1$ for $x \ll 1$ and $g(x) \sim x^\psi$ for $x \gg 1$. As
mentioned in Ref.~\cite{bustingorry_thermal_rounding_epl} strong
corrections-to-scaling effects are present in these results. In order to show
that, Fig.~\ref{f:vdeT-scal-Ap05-n} presents an attempt to use the standard
finite size scaling correction scaling, Eq.~\eqref{eq:std-fsc}, with the bare
data in Fig.~\ref{f:vdeT-scal-Ap05-g-a}. One can observe strong finite size
corrections and this can also be observed with other values of $R_0$.
In addition, the collapse of the data does not improve significantly when
using other values of the scaling exponents. Despite
these strong finite-size effects exhibited by the velocity at critical force,
the power-law regime characterized by the thermal rounding exponent $\psi=0.15$
does not suffer from strong finite size effects, as shown in the following
sections.

\section{Structure factor analysis}
\label{sec:disorder}

In this section we turn to the complementary geometrical
analysis of the structure factor, which contains
information on the geometry of the string at different length scales. The
results presented in this section complements those reported in
Refs.~\cite{bustingorry_thermal_rounding_epl,bustingorry_periodic} by including
different disorder strengths.

From the numerical simulations, the steady state structure factor is defined as
\begin{equation}
 S_q = \frac{1}{L}\overline{\left\langle \left| \sum_{j=0}^{L-1} u_j\, e^{iqj}
\right|^2 \right\rangle},
\end{equation}
where $q=2 \pi n/L$, with $n = 1,...,L-1$. One can show using dimensional
analysis that when the width $w$ of a self-affine interface of size $L$ is
described through a roughness exponent $\zeta$, i.e. $w \sim L^\zeta$,
then the structure factor behaves as $S_q \sim q^{-(1+2\zeta)}$ in $1+1$
dimensions.

\begin{figure}[!tbp]
\includegraphics[angle=-0,width=7.5cm,clip=true]{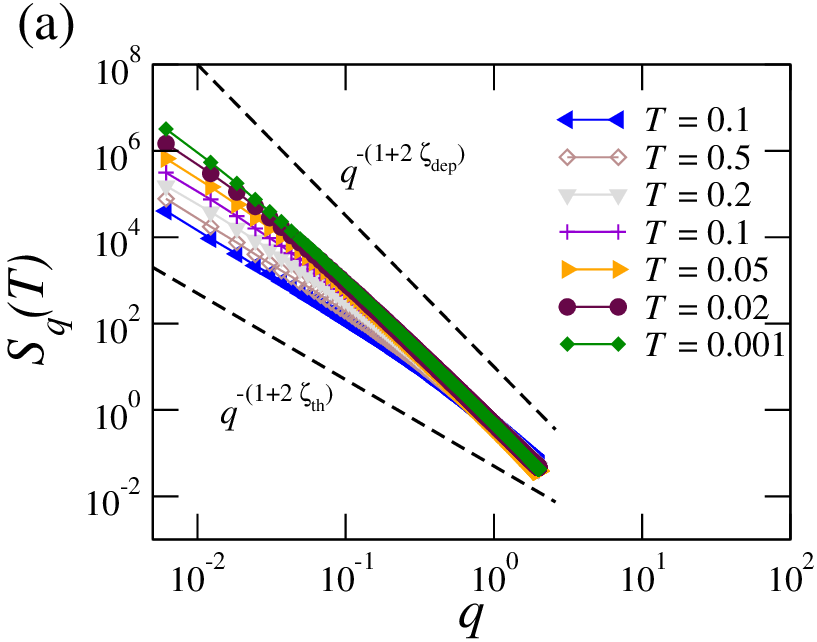}
\includegraphics[angle=-0,width=7.5cm,clip=true]{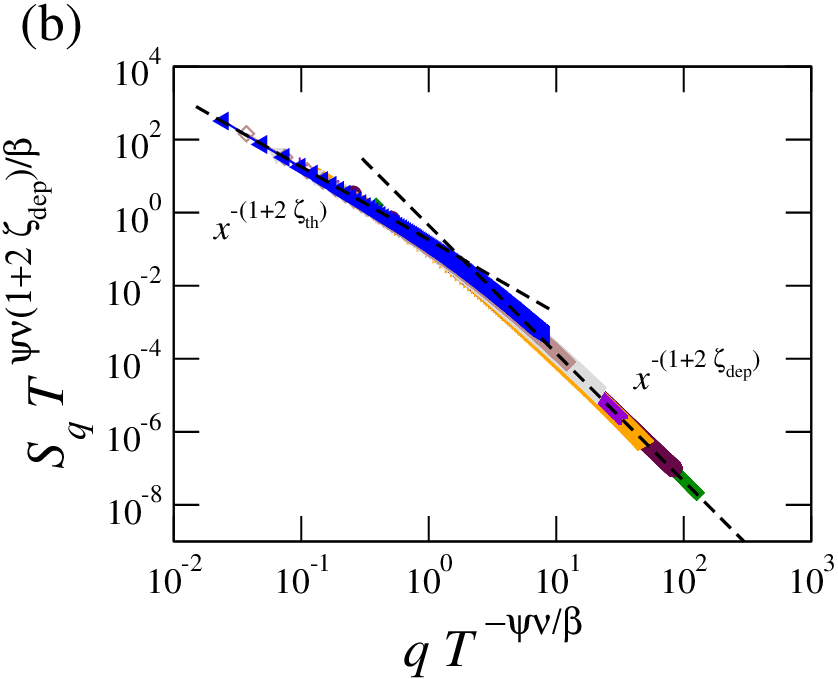}
\caption{\label{f:Sdeq-log-des-05a} (Color online) (Color online) Structure factor and its
scaling function for $R_0=0.5$. The curves in (a) correspond to the critical
force and different temperatures. The size of the sample is given by $L=1024$
and $M=4L^{\zeta_{\mathrm{dep}}}$. (b) Scaled curves showing the crossover
between the depinning regime at small length scales ($x=qT^{-\psi \nu/\beta} \gg 1$) and the thermal regime at
large length scales ($x=qT^{-\psi \nu/\beta} \ll 1$). Together with $\psi=0.15$, the values
$\beta=0.33$~\cite{duemmer2}, $\nu=1.33$~\cite{kolton_short_time_exponents} and
$\zeta_{\mathrm{dep}}=1.25$~\cite{duemmer2,rosso_hartmann} were also used.
}
\end{figure}

\begin{figure}[!tbp]
\includegraphics[angle=-0,width=7.5cm,clip=true]{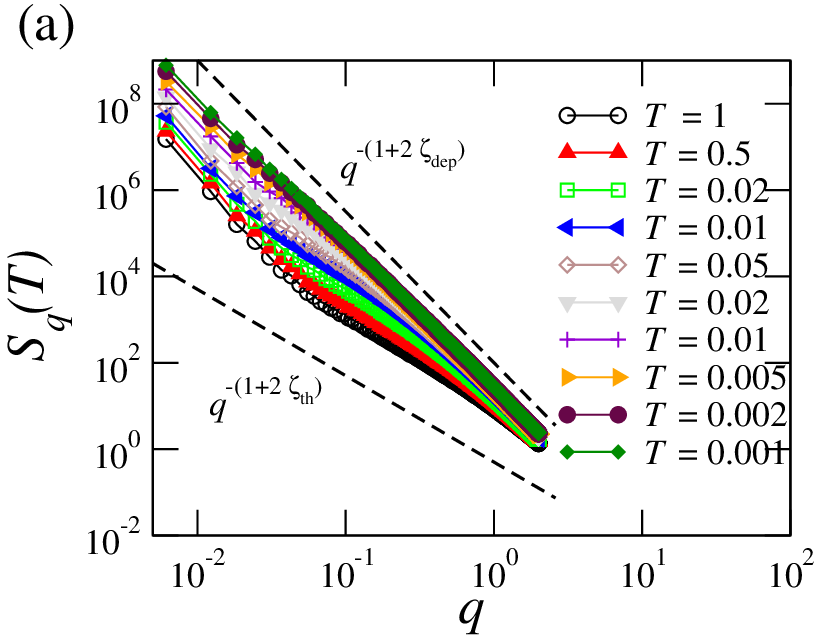}
\includegraphics[angle=-0,width=7.5cm,clip=true]{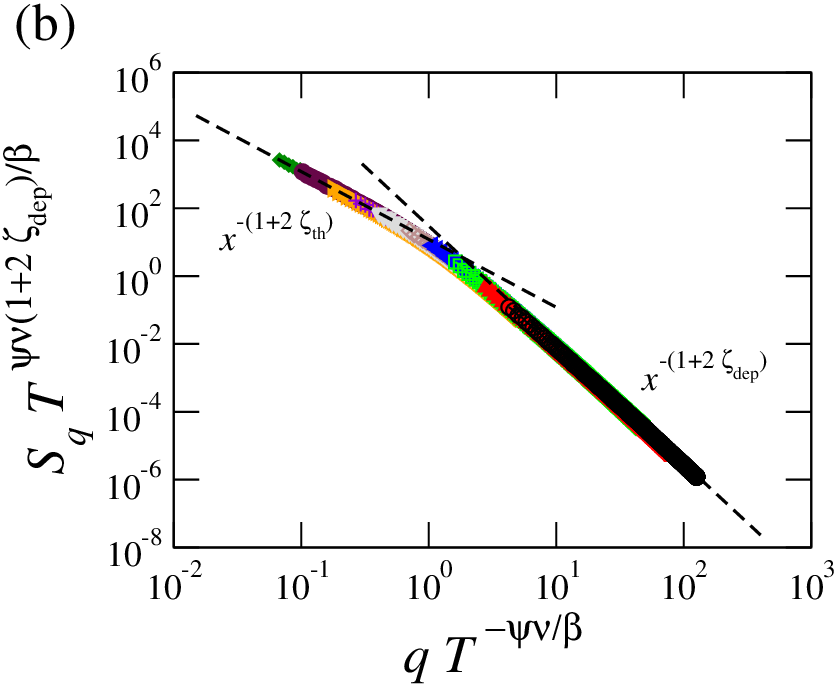}
\caption{\label{f:Sdeq-log-des-5a} Structure factor and its
scaling function for $R_0=5$. The curves in (a) correspond to the critical force
and different temperatures. The size of the sample is given by $L=1024$ and
$M=L^{\zeta_{\mathrm{dep}}}$. (b) Scaled curves showing the crossover between
the depinning regime at small length scales ($x=qT^{-\psi \nu/\beta} \gg 1$) and the thermal regime at large
length scales ($x=qT^{-\psi \nu/\beta} \ll 1$). The very large length scale random-periodic fast-flow regime has
been discarded (see the text). Together with $\psi=0.15$, the values
$\beta=0.33$~\cite{duemmer2}, $\nu=1.33$~\cite{kolton_short_time_exponents} and
$\zeta_{\mathrm{dep}}=1.25$~\cite{duemmer2,rosso_hartmann} were also used.
}
\end{figure}

At small length scales, $q\gg 1/\xi$, the structure factor shows the typical
roughness regime associated to depinning, i.e. $S_q \sim q^{-(1+2
\zeta_{\mathrm{dep}})}$, while at large length scales, $q \ll 1/\xi$,
fluctuations are dictated by effective thermal fluctuations induced by the
disorder, i.e. $S_q \sim q^{-(1+2 \zeta_{\mathrm{th}})}$. The thermal and
depinning roughness exponents are, respectively, $\zeta_{\mathrm{th}} = 1/2$ and
$\zeta_{\mathrm{dep}} = 1.25$~\cite{rosso_hartmann,duemmer2}. In the critical
region the depinning correlation length is given by the velocity as $\xi \sim
v^{-\nu/\beta}$. Thus, the depinning correlation length depends on the
temperature only through the velocity and in the thermal rounding regime $\xi
\sim T^{-\psi \nu/\beta}$~\cite{bustingorry_thermal_rounding_epl}. With this
information one can write for the structure factor that
\begin{equation}
 \label{eq:scal-Sq}
 S_q = T^{-\psi \nu (1+2 \zeta_{\mathrm{dep}})/\beta} s\left( qT^{-\psi
\nu/\beta} \right),
\end{equation}
where the scaling function $s(x) \sim x^{-(1+2\zeta_{\mathrm{th}})}$ for $x \ll
1$ and $s(x) \sim x^{-(1+2\zeta_{\mathrm{dep}})}$ for $x \gg 1$. In
Ref.~\cite{bustingorry_thermal_rounding_epl} we showed that the structure factor
scales with the previous form using $L=1024$ and $M=L^{\zeta_{\mathrm{dep}}}$
for the disorder intensity $R_0 =1$. Here, we show in
Fig.~\ref{f:Sdeq-log-des-05a}(a) the temperature dependence of the structure
factor corresponding to $R_0 = 0.5$, $L=1024$ and $M=4L^{\zeta_{\mathrm{dep}}}$.
For these parameters the presented data do not show transverse finite size
effects~\cite{bustingorry_periodic}. Figure~\ref{f:Sdeq-log-des-05a}(b) shows
the scaling of the structure factor according to Eq.~\eqref{eq:scal-Sq} and
using $\psi = 0.15$, which shows a very satisfactory data collapse.

In order to reach the steady-state for the same temperatures of
Fig.~\ref{f:Sdeq-log-des-05a}(a) but larger disorder intensities it is necessary
to equilibrate the system for longer times. Since this equilibration time scales
with the transverse system size we can reduce the simulation time by using
$M=L^{\zeta_{\mathrm{dep}}}$ for $R_0=5$. The resulting data, shown in
Fig.~\ref{f:Sdeq-log-des-5a}(a), presents the small length scale depinning
regime and the large scale effective thermal regime described above, but also
clearly show a larger length scale regime where finite transverse size effects
are present. In this regime the roughness exponent is the one corresponding to a
random-periodic system in the fast-flow regime, $\zeta_{\mathrm{per}} =
3/2$~\cite{bustingorry_periodic}. Hence, we can detect and discard the data
corresponding to this random-periodic regime in order to get a curve that can be
scaled again using Eq.~\eqref{eq:scal-Sq} and the known random-manifold
exponents, as shown in Fig.~\ref{f:Sdeq-log-des-5a}(b), getting again a very
satisfactory collapse.

Therefore, we have presented here data of the structure factor for different
disorder intensities which shows that the quoted thermal rounding exponent is
disorder independent. Furthermore, we have shown how the thermal rounding
exponent gives the temperature dependence of the depinning correlation length,
$\xi \sim T^{-\psi \nu/\beta}$, from a steady-state geometry analysis.

\section{Short-time dynamics analysis}
\label{sec:short-time}

\begin{figure}[!tbp]
\includegraphics[angle=-0,width=7.5cm,clip=true]{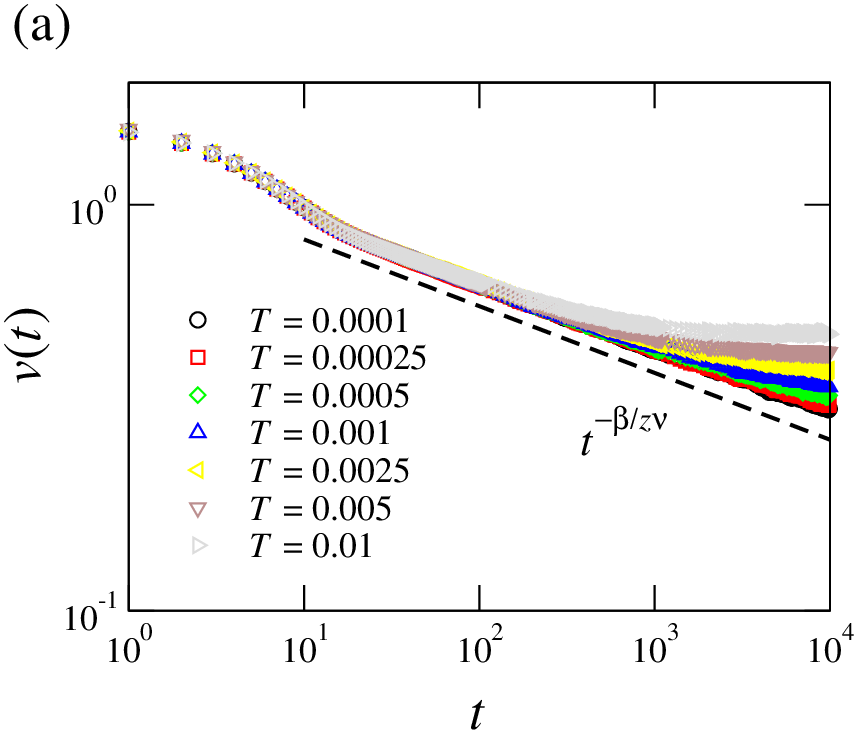}
\includegraphics[angle=-0,width=7.5cm,clip=true]{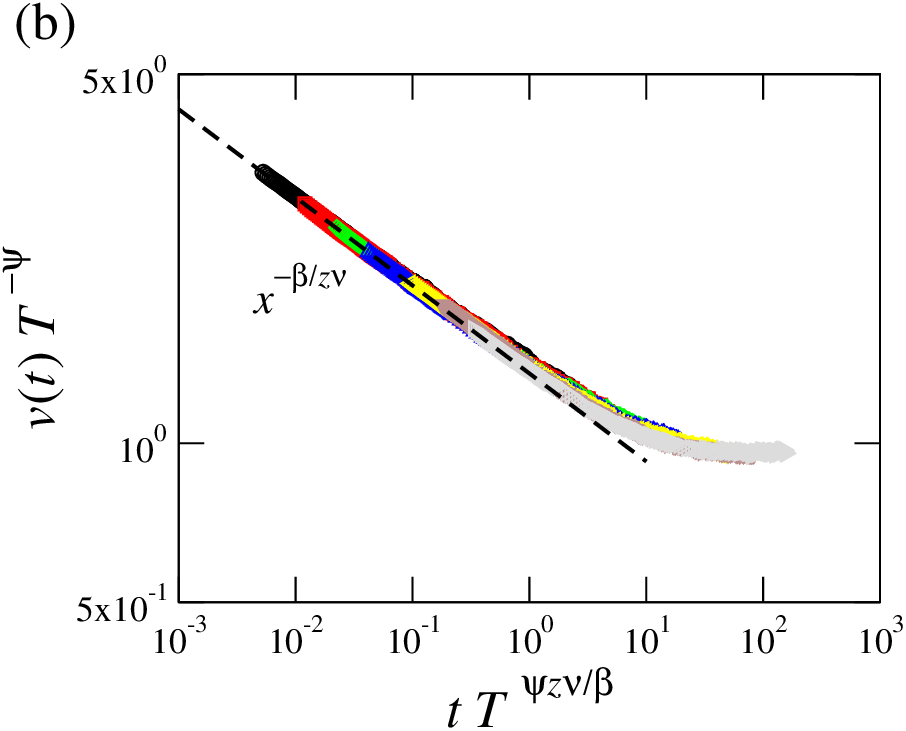}
\caption{\label{f:vdet_sht} (Color online) (a) Short time evolution of the velocity at the
critical force and for different temperatures, from $T=0.0001$ (lower curve) to $T=0.01$
(upper curve), as indicated in the key. Data correspond to
$R_0 = 1$, $L = 1024$ and $M = L^{\zeta_{\mathrm{dep}}}$.(b) Short
time scaling of the velocity for $t>20$. The behavior $x^{-\beta/z \nu}$ for 
$x=t T^{\psi z \nu/\beta}\ll1$ is indicated with a dashed line. Together with $\psi=0.15$, the values
$\beta=0.33$~\cite{duemmer2}, $\nu=1.33$~\cite{kolton_short_time_exponents} and
$z = 3/2$~\cite{kolton_short_time_exponents} were also used.}
\end{figure}

One possible way to get rid of finite size effects is to analyze the short time
dynamics. Starting from a given non-steady initial condition at fixed force $F$
and temperature $T$, the velocity begins to evolve with time until it reaches
the steady-state value corresponding to the values of $F$ and $T$. This
transient dynamics is controlled, at short times, by a single growing
correlation length, $\xi(t)$, which at longer times saturates to the
steady-state correlation length above threshold, $\xi \sim v^{-\nu/\beta}$.
Since the transient correlation length grows as $\xi(t) \sim t^{1/z}$, with
$z \approx 3/2$~\cite{kolton_short_time_exponents} the depinning dynamical exponent,
scaling arguments show that the velocity decreases with time as $v(t) \sim
\xi(t)^{-\beta/\nu} \sim t^{-\beta/z\nu}$~\cite{kolton_short_time_exponents}
before saturating to the steady-state value above threshold, given by $v(t \to
\infty) \sim f^\beta$ at $T=0$ or by  
$v(t \to \infty) \sim T^\psi$ at $f=0$.

Figure~\ref{f:vdet_sht}(a) shows the time evolution of the velocity exactly at
the critical force and for different temperature values as indicated. The dashed
line corresponds to the expected short-time critical behavior $v(t) \sim
t^{-\beta/z\nu}$. Discarding the very short time regime, $t \le 20$, which
contains information about the microscopic non-universal
dynamics~\cite{kolton_short_time_exponents}, the curves in
Fig.~\ref{f:vdet_sht}(a) can be recast into a universal form using the scaling
function
\begin{equation}
 v(t) = T^{\psi} h\left( t T^{\psi z \nu/\beta} \right),
\end{equation}
with $h(x) \sim x^{-\beta / z \nu}$ for $x \ll 1$ and $h(x) \sim 1$ for $x \gg
1$. The data collapse shown in Fig.~\ref{f:vdet_sht}(b) uses the previously
known depinning exponents $\beta=0.33$~\cite{duemmer2},
$\nu=1.33$~\cite{kolton_short_time_exponents},
$z = 3/2$~\cite{kolton_short_time_exponents}, together with
the thermal rounding exponent $\psi = 0.15$. Since the data collapse is good
with no need of adjustable parameters we can conclude that the value of the
thermal rounding exponent obtained in
Ref.~\cite{bustingorry_thermal_rounding_epl} is consistent and does not suffer
from strong finite-size effects.

\section{Velocity scaling function around depinning}
\label{sec:univ-func}

\begin{figure}[!tbp]
\includegraphics[angle=-0,width=7.5cm,clip=true]{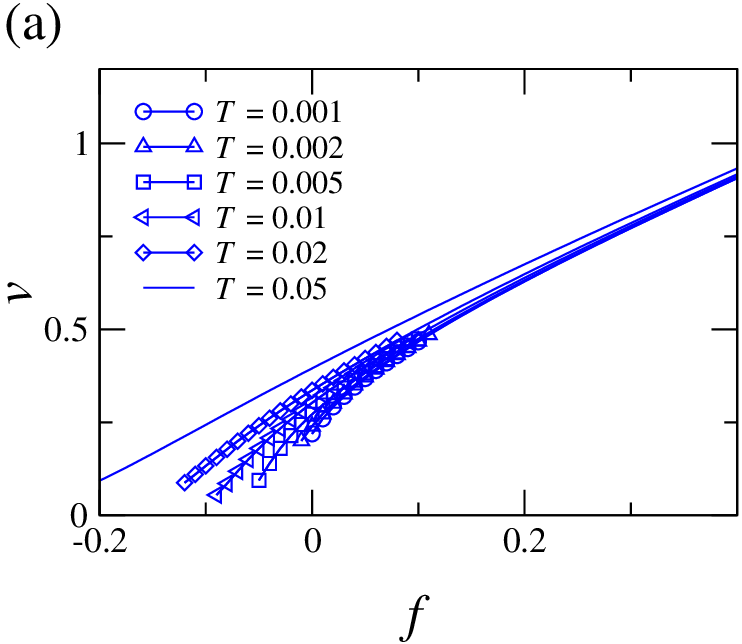}
\includegraphics[angle=-0,width=7.5cm,clip=true]{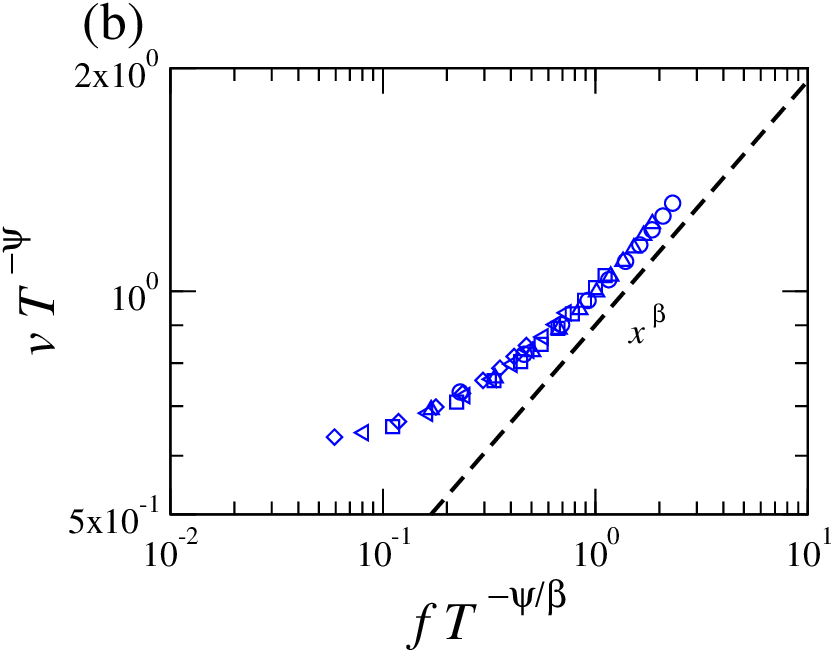}
\caption{\label{f:vf_temp_1} (Color online) (a) Velocity-force characteristics at finite
temperatures for $R_0 =1$. Data shown with points (lines) are inside (outside)
the thermal rounding region (see the text). (b) Scaling of the data for $f>0$
and in the thermal rounding region using the scaling form given in
Eq.~\eqref{eq:universalv}. The dashed line indicates the asymptotic expected form
$x^\beta$ for $x=f T^{-\psi/\beta} \gg 1$. Here, the value $\beta=0.33$~\cite{duemmer2} was used together with
$\psi=0.15$.}
\end{figure}

\begin{figure}[!tbp]
\includegraphics[angle=-0,width=7.5cm,clip=true]{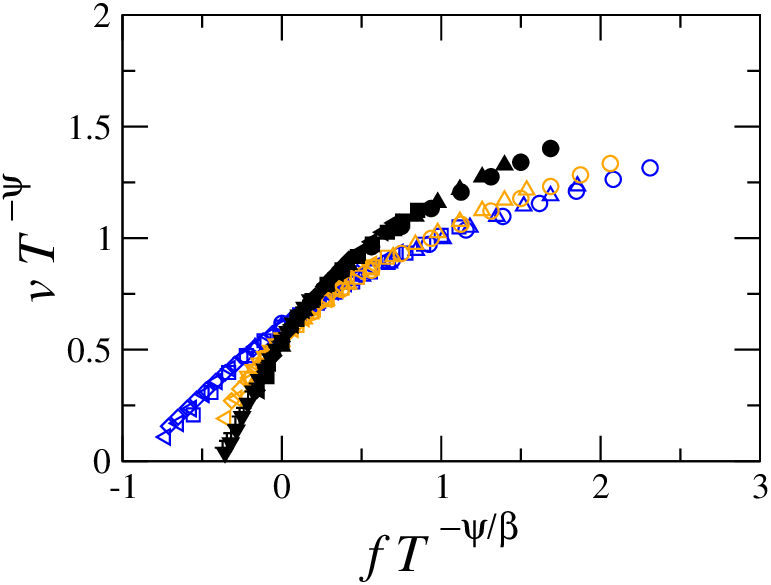}
\caption{\label{f:vf_temp_des} (Color online) Scaling of velocity-force curves for different
temperatures and close to the thermal rounding regime. Different disorder
intensities are shown: $R_0=1$ (blue), $R_0=2$ (orange) and $R_0=5$ (black).
Symbols represent different temperatures as in Fig.~\ref{f:vf_temp_1}(a). Here,
the value $\beta=0.33$~\cite{duemmer2} was used together with $\psi=0.15$.}
\end{figure}

In this section we turn to the analysis of the universal behavior of the force
and temperature dependent velocity function, $v(f,T)$. Focusing on testing the robustness
of the thermal rounding exponent $\psi$, parameter values around the critical
point given by $f=0$ and $T=0$ are tested. If there were not strong finite size
effects, in the vicinity of the critical region the velocity should scale as
\begin{equation}
\label{eq:universalv}
 v T^{-\psi} \sim H\left( f T^{-\psi/\beta} \right),
\end{equation}
with $H(x) \sim x^\beta$ for $x\gg 1$. Figure~\ref{f:vf_temp_1}(a) shows
velocity-force curves for different temperatures and for $R_0=1$. The numerical
data is split into two sets: on the one hand data points correspond to given
parameters which are ``inside'' the thermal rounding region, and on the other
hand continuous lines represent data ``outside'' the thermal rounding region.
The data are outside the critical region either because temperature is too high,
$T>0.02$ in the present case, or because the force is far away from the critical
value, $|f| \gg 1$ ($f \gg 1$ and $f \ll -1$ corresponding to the fast-flow and
creep regimes, respectively). In addition, to avoid finite-size effects, data 
points are also considered ``outside'' the critical thermal rounding region
if they correspond to velocities smaller than the crossover 
at $v \sim v_{min}$ to single-particle behavior for each size $L$. Since in the critical region
$\xi \sim v(f,T)^{-\nu/\beta}$, as shown from the structure factor analysis, 
we roughly have $v_{min} \sim L^{-\beta/\nu}$. 
According to such criteria, the selected data is finally presented in the
scaled form Eq.~\eqref{eq:universalv} in Fig.~\ref{f:vf_temp_1}(b) for $f>0$. The
dashed line indicates the expected asymptotic $x^\beta$ form, corresponding to
the scaling function $H$ around the critical region. The collapse into a single 
curve for different $T$ and $f$ confirms numerically that the data set used 
is inside the critical scaling region.

\begin{figure}[!tbp]
\includegraphics[angle=-0,width=7.5cm,clip=true]{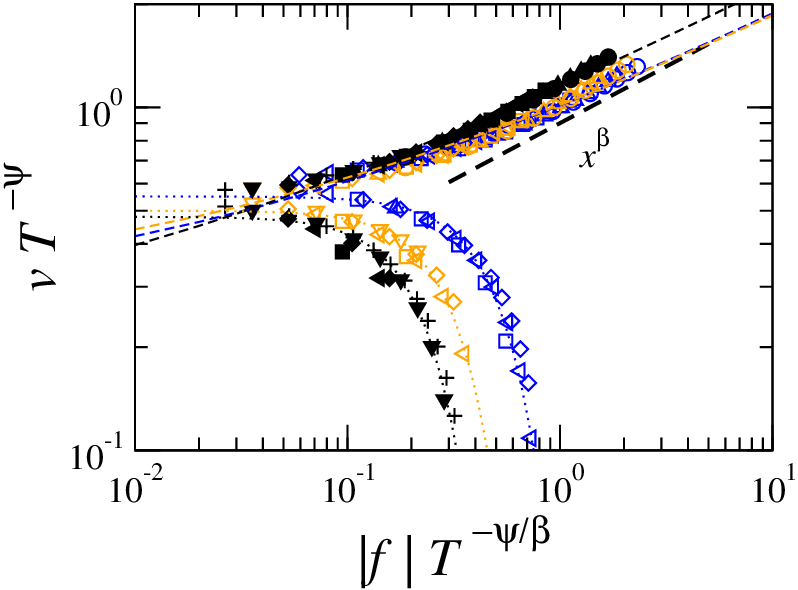}
\caption{\label{f:vf_temp_des-fit} (Color online) Same data as in Fig.~\ref{f:vf_temp_des} but
in double-logarithmic representation. The upper and lower curves correspond to
$f>0$ and $f<0$, respectively. We also show fitting curves using
$a_1+a_2x^\beta$ and $b_1\exp[-b_2|x|^{\beta/\psi}]$, as suggested by the
scaling functions Eqs.~\eqref{eq:VFT-reduced-} and \eqref{eq:VFT-reduced+}.}
\end{figure}

The scaling function $H$ is not yet universal as it also depends on the disorder intensity.
Figure~\ref{f:vf_temp_1} shows the critical region and the form of $H$ for
$R_0=1$. In Fig.~\ref{f:vf_temp_des} we show the velocity scaling function $H$
for different disorder intensities, $R_0=1,2$ and $5$, for the full force range
within the thermal rounding region. In Fig.~\ref{f:vf_temp_des-fit} the same
data is presented in a double logarithmic scale. As can be observed all curves
display the asymptotic power-law form $H \sim x^\beta$ for $f \gg
T^{\psi/\beta}$, but with different prefactors for each disorder intensity.

At this point, in order to properly include the disorder intensity on the
scaling velocity function and obtain the universal function, 
a disorder-dependent temperature scale $T_c$ is needed.
Again, for the simple example of the depinning of a particle in a periodic potential 
$U(x)=R_0 \cos(x/\lambda)$, 
$\gamma dx/dt = -dU(x)/dx + F + \eta(t)$, with $\eta(t)$ a Langevin noise at 
temperature $T$, it is easy to see, from pure dimensional analysis, 
that $T_c \sim R_0$, and therefore $\gamma V/F_c = h[(F-F_c)/F_c, T/R_0]$,  
where $F_c = R_0/\lambda$. 
Such naive approach does not work for the elastic string, as the characteristic energy scale 
is not simply $R_0$ as for the particle, 
but it arises from the interplay of disorder and elasticity.
Although it is not obvious that it should work at depinning, one is tempted to
use the scaled temperature $\tau=T/T_c$, where $T_c=U_c/k_B$ gives the
characteristic energy scale in the creep regime at small
forces~\cite{ioffe_creep,nattermann_creep}. This energy scale is however 
not universal and depends on microscopic details of the disorder
~\cite{ioffe_creep,nattermann_creep,nattermann_hysteresis_domainwall}.
The assessment of the full dependence of $U_c$ on microscopic parameters is
thus not straightforward and from a pragmatic point of view 
one could directly fit it from the creep law. As shown
with numerical simulations within the creep regime, at larger temperatures than
the one used here, $U_c$ can be fitted from the creep law, but it dependence on
$R_0$ is not trivial~\cite{kolton_creep2}. We do not have access to $T_c$ with
the present numerical results, which focused in the force region around the
critical depinning threshold. 

Therefore, we can not incorporate at this stage the
influence of the disorder intensity in the velocity function. In spite of this, 
the data displayed in Fig.~\ref{f:vf_temp_des} clearly show that the velocity can be
represented in a scaled form, with identical critical exponents, 
for different disorder intensities. More importantly, these data supports 
the disorder independent value $\psi=0.15$ tested here.

Finally, it is worth relating our results with the universal scaling function
proposed by Nattermann, Pokrovsky and
Vinokur~\cite{nattermann_hysteresis_domainwall} using a phenomenological interpolating 
form for the full force and temperature dependence of the velocity of a domain wall
in a random medium. This form includes the thermal rounding regime around $F_c$
and the $F \ll F_c$ creep regime, thus depending also on the universal creep exponent
$\mu=1/4$ (for a one dimensional elastic interface). The proposed functional form in
Ref.~\cite{nattermann_hysteresis_domainwall} is different below and above the
critical force and can be written as
\begin{equation}
\label{eq:VFT-below}
 V = m F \frac{\exp \left[ -\frac{T_c}{T} \left( 1-\frac{F}{F_c}
\right)^{\beta/\psi} \left( \frac{F_c}{F} \right)^\mu \right]}{1+\left[
\frac{T_c}{T} \left( \frac{F_c}{F} \right)^\mu \right]^\psi},
\end{equation}
for $F<F_c$ and
\begin{equation}
 \label{eq:VFT-above}
 V = \frac{m F}{1+\left[ \frac{T_c}{T} \left( \frac{F_c}{F} \right)^\mu
\right]^\psi} + m F \left( 1-\frac{F_c}{F} \right)^\beta,
\end{equation}
for $F>F_c$. It can be shown that close to the depinning region, i.e. above the
creep regime, where $f = (F-F_c)/F_c \ll 1$ and $\tau = T/T_c \ll 1$ this
phenomenological form can be reduced to the scaling form
Eq.~\eqref{eq:universalv}, with $H(x) = H^-(x)$ for $f<0$ and $H(x) = H^+(x)$ for
$f>0$. The corresponding limit functions are
\begin{eqnarray}
\label{eq:VFT-reduced-}
 H^-\left( f \tau^{-\psi/\beta} \right) = e^{-(-f\tau^{-\psi/\beta})^{\beta/\psi}}, \\
\label{eq:VFT-reduced+}
 H^+\left( f \tau^{-\psi/\beta} \right) = 1+(f \tau^{-\psi/\beta})^\beta.
\end{eqnarray}

Since we do not have the temperature scale $T_c$ from the creep law, we have
directly fitted the data for the velocity scaling function to the universal
forms suggested by Eqs.~\eqref{eq:VFT-reduced-} and \eqref{eq:VFT-reduced+}. The $f<0$ and $f>0$ ranges have
been fitted separately using $a_1+a_2x^\beta$ and
$b_1\exp[-b_2|x|^{\beta/\psi}]$, respectively, obtaining four fitting parameters
for each disorder intensity. The results are shown in
Fig.~\ref{f:vf_temp_des-fit}. In all cases the fit is better in the $f<0$
region. Furthermore, one can observe that the obtained curves interpolate badly
around $f=0$. In fact, enforcing $a_1=b_1$ makes the fitting considerably worse.
We therefore conclude that the data can not be satisfactorily fitted using this
phenomenological form, particularly above threshold, hence evidently pointing to
the need of a more accurate description of the thermal rounding of the depinning
transition.

The phenomenological functional forms, Eqs.~\eqref{eq:VFT-below} and
~\eqref{eq:VFT-above}, give a potentially important tool which allows to
directly fit experimental data. This was directly used in
Ref.~\cite{metaxas_thesis}, where the velocity-force characteristic below
threshold for ultrathin ferromagnetic layers was fitted using
Eq.~\eqref{eq:VFT-below}. By fitting just one experimental curve below
threshold, the value $\psi = 0.15 \pm 0.10$ was obtained for the thermal
rounding exponent. Since several fitting parameters were used and due to the
large error bar, this value can only be compared with our numerical value with
extreme caution. Anyway, the experimental value is consistent with our numerical
simulations.

\section{Summary}
\label{sec:conc}

We have presented extensive numerical simulations to test the validity of the
thermal rounding exponent of the depinning transition. We analyzed 
the direct scaling of the steady-state velocity-force characteristics, the steady-state 
structure factor and the short-time transient dynamics. The existence 
of a critical (power law) thermal rounding of
the depinning transition is consistent with all our results, together with the 
existence of a unique divergent length scale, dependent on temperature and/or distance to
the critical pinning force, but ultimately controlled by the velocity as in 
the zero temperature depinning transition. The results 
are all consistent with a value of the thermal rounding exponent of $\psi = 0.15$ 
in agreement with our previously
reported value~\cite{bustingorry_thermal_rounding_epl}. This exponent describes
the power-law vanishing of the velocity with temperature exactly at the critical
depinning force, $V \sim T^\psi$, for the universality class 
of one dimensional elastic interfaces with short-range elasticity 
and short-range correlations in the disorder.

Although the value of the thermal rounding exponent have been previously
obtained with larger system sizes, where finite size corrections are still
observable, we have shown here that this value is also consistent with
short-time dynamics results which do not suffer from severe finite size effects.
Besides, $\psi = 0.15$ also describes the scaling properties of the structure
factor for various disorder strength values, connecting this value with a
geometrical roughness crossover in the interface. Finally, we have shown that it
is consistent with a scaling function describing the velocity-force
characteristics as a function of temperature and force. Experimental
confirmation of our results, directly targeting the thermal rounding 
regime and allowing to test the value of the thermal rounding exponent, 
would be welcome.

\begin{acknowledgments}
This work was supported in part by the Swiss National Science Foundation under
MaNEP and Division II. SB and ABK acknowledge financial support from ANPCyT
Grant No. PICT2007886 and CONICET Grant No. PIP11220090100051.
ABK acknowledges Universidad de Barcelona, Ministerio de Ciencia e
Innovaci\'on (Spain) and Generalitat de Catalunya for partial support
through I3 program.
\end{acknowledgments}

\bibliography{tfinita4}

\end{document}